\documentclass[twocolumn,twocolappendix]{aastex631}

\usepackage{xspace}
\usepackage{amsmath}
\usepackage{enumerate}

\newcommand\Abacus{\textsc{Abacus}\xspace}
\newcommand\Mnl{\ensuremath{M_\mathrm{nl}}\xspace}
\newcommand\Rnl{\ensuremath{R_\mathrm{nl}}\xspace}
\newcommand\knl{\ensuremath{k_\mathrm{nl}}\xspace}
\newcommand{\code}[1]{\texttt{#1}}

\defcitealias{Joyce+2021}{J2021}
\defcitealias{Garrison+2021b}{G2021b}

\shorttitle{Scale-free $k$NN}
\shortauthors{Garrison et al.}
\graphicspath{{./}{figures/}}

\begin{document}

\title{Self-Similarity of $k$-Nearest Neighbor Distributions in Scale-Free Simulations}

\author[0000-0002-9853-5673]{Lehman H. Garrison}
\affiliation{Center for Computational Astrophysics, Flatiron Institute, 162 5\textsuperscript{th} Avenue, New York, NY 10010, USA}

\author[0000-0002-5969-1251]{Tom Abel}
\affiliation{Kavli Institute for Particle Astrophysics and Cosmology, Stanford University, 452 Lomita Mall, Stanford, CA 94305, USA} 
\affiliation{Department of Physics, Stanford University, 382 Via Pueblo Mall, Stanford, CA 94305, USA}
\affiliation{SLAC National Accelerator Laboratory, 2575 Sand Hill Road, Menlo Park, CA 94025, USA}

\author{Daniel J. Eisenstein}
\affiliation{Center for Astrophysics $|$ Harvard \& Smithsonian
60 Garden Street, Cambridge, MA 02138, USA}



\begin{abstract}
We use the $k$-nearest neighbor probability distribution function ($k$NN-PDF, Banerjee \& Abel 2021) to assess convergence in a scale-free $N$-body simulation.  Compared to our previous two-point analysis, the $k$NN-PDF allows us to quantify our results in the language of halos and numbers of particles, while also incorporating non-Gaussian information.  We find good convergence for 32 particles and greater at densities typical of halos, while 16 particles and fewer appears unconverged.  Halving the softening length extends convergence to higher densities, but not to fewer particles.  Our analysis is less sensitive to voids, but we analyze a limited range of underdensities and find evidence for convergence at 16 particles and greater even in sparse voids.
\end{abstract}


\section{Introduction} \label{sec:intro}
Cosmological $N$-body simulations model the evolution of the continuous Vlasov-Poisson distribution function with a set of $N$ particles interacting under mutual gravitational attraction.  The discrete nature of the particles imposes a small-scale cutoff to the resolution of the simulation, although precisely the length and mass scale of such a cutoff and the mechanism by which it operates is not so clear.  In some limits, such as the weakly perturbative regime, it is possible to show that the correspondence of the dynamics to the continuum solution quickly degrades as one approaches the mean interparticle separation, $\ell = N^{1/3}/L$ \citep[for box size $L$,][]{Garrison+2016}.  However, in the strongly non-linear regime, resolution is often obtained at scales many times smaller than the mean particle separation.  The relevant cutoff is therefore not $\ell$, but something less obvious, and the question becomes the scale of such a cutoff and the mechanism by which it manifests.  And to complicate matters, additional parameters, such as softening, are often introduced to regularize the particle interactions.  Their role in setting the resolution limit of a simulation must also be assessed.

Traditional convergence tests operate by tuning a discreteness parameter, such as $N$, towards its continuum value, in this case infinity.  But such tests are limited in their dynamic range, as the computational expense necessarily increases towards the continuum.  Furthermore, parameters such as the softening length $\epsilon$ do not have a well-defined continuum value---consider that taking $\epsilon\rightarrow0$ will increase scattering encounters between particles, which does not occur in the continuous Vlasov-Poisson system.

Another class of tests exists that exploits the self-similar nature of gravity, known as scale-free simulations.  With a power-law power spectrum of index $n_s$ and an $\Omega_M=1$ background cosmology, only one scale is present in the continuum version of the problem, with time and space related by a power law whose index is determined by $n_s$.  In other words, a small-scale property of the simulation at one time must be equal to the same property measured on large-scales at a later time.  Where such self-similarity is not observed, the simulation may be said to differ from the continuum solution.

Scale-free simulations have a rich history in the $N$-body literature \citep[e.g.][]{Efstathiou+1988,Colombi+1996,Jain+1998,1999ApJ...520...35S,Smith+2003,Widrow+2009,Orban_Weinberg_2011}, alternately being used to build $\Lambda$CDM predictions and as tests of $N$-body dynamics.  Scale-free tests do not establish the absolute convergence of a simulation, but do guarantee that any remaining errors must themselves be of a self-similar nature.  For many classes of error, this is sufficient, as UV cutoffs like $N$ and $\epsilon$, as well as IR cutoffs like $L$, impose a preferred scale.  A notable exception is certain classes of time stepping; for example, a fixed time step in log-$a$ will yield self-similar results for any time step size, even implausibly large.  Even time step schemes that are not log-constant, such as \Abacus's, seem to exhibit moderately self-similar errors, as discussed in \cite[hereafter \citetalias{Joyce+2021}]{Joyce+2021}.  Such errors can be controlled with more traditional tests of the raw amplitude of clustering or other non-rescaled statistics.

In this work, we seek to build upon our previous scale-free tests, which were limited to two-point correlations \citep[2PCF;][hereafter \citetalias{Garrison+2021b}]{Joyce+2021,Maleubre+2021,Garrison+2021b}.  This is both a result of computational expediency---the 2PCF and power spectrum are more readily computed than the 3PCF or bispectrum---and because two-point functions encode the full information content of Gaussian random fields.  As these simulations are imparted with Gaussian initial conditions, two-point statistics are an excellent summary of information contained in large scales, so long as the density fluctuations on such scales remain perturbatively small.  However, as many of the issues we wish to explore relate to the deeply non-linear, small-scale behavior of $N$-body simulations---where the density field is certainly not Gaussian---we must assume that the 2PCF is an incomplete compression the information available.

In this work, we turn to the $k$-nearest neighbor probability distribution function ($k$NN-PDF) as a non-Gaussian summary statistic.  The $k$NN-PDF gives the distance distribution to the $k$-th nearest neighbor from a random point in the volume, and is connected to all higher $N$-point functions.  Furthermore, as it is defined on sets of discrete points, it allows us to quote our results in terms of numbers of particles, an advantage not afforded by two-point methods.  Finally, its computational expense is modest compared to direct evaluation of 3-point and higher statistics.  This is particularly important for measuring statistics directly on $N$-body particles, as the number density is orders of magnitude higher than that of halos or galaxies.

This paper is organized as follows.  In Section~\ref{sec:definitions}, we define the nearest-neighbor distribution, scale-free simulations, and the mapping by which the $k$NN-PDF may be compared self-similarly.  In Section~\ref{sec:results}, we present measurements of the $k$NN-PDF on an $n=-2$ $N$-body simulation and assess their self-similarity in the halo and void regimes.  In Section~\ref{sec:conclusions}, we summarize and conclude.

\section{$\lowercase{k}$-Nearest Neighbor Distributions and Scale-Free Simulations}\label{sec:definitions}
\subsection{Nearest Neighbor Distributions}
The $k$-nearest neighbor probability distribution function ($k$NN-PDF) is a measure of the spatial clustering of a set of discrete points.  It quantifies the probability density of the $k$-th nearest point lying at distance $r$ from a random point in space.  Introduced in \cite{Banerjee_Abel_2021,Banerjee_Abel_2021b} for measurement of cosmological clustering, it is connected to the counts-in-cells, void probability function, and all $N$-point correlation functions of the density field.  Furthermore, it is computationally inexpensive to compute, and makes for an efficient compression of the information---both Gaussian and non-Gaussian---contained within the field.  It has relatively few free parameters: $k$, and any parameters used in estimating the PDF, such as the histogram bin width.

The $k$NN-PDF may be computed as follows.  First, the points are organized into a space-partitioning data structure that allows for efficient nearest-neighbor queries, such as a grid or a tree.  In this work, we employ \textsc{SciPy}'s \code{KDTree} \citep{Bentley_1975,SciPy_2020}.  Then, a set of $N_R$ random points are generated uniformly in the volume, and the tree is queried for each, returning the distance to the $k$-th nearest data point for each random point.  Then, the $N_R$ distances are histogrammed and normalized to form the PDF, or, if the cumulative distribution function is desired, the list is sorted.

The $k$NN-PDF offers a different view on the density field than the 2PCF.  As already discussed, it is a non-Gaussian summary statistic, containing information from all $N$-point functions.  But additionally, it is a volume-weighted statistic, in that the random points are uniformly distributed in space.  Each random point gets one ``vote'' in the $k$NN-PDF; high-density and low-density regions are weighted equally.  Compare with the small-scale 2PCF, which is dominated by dense regions because the statistic is pair-weighted.  The $k$NN-PDF, by contrast, probes high-density regions in its small-$r$ tail and low-density regions in its high-$r$ tail.

Finally, the $k$NN-PDF seems particularly well-suited to studies of particle systems like halos, as it operates on sets of points rather than continuous fields.  Because $k$ is an input to the algorithm, it is automatically spatially adaptive in the sense that it will return the size of the spheres containing $k-1$ particles, no matter the radius.  This allows mapping convergence not just with respect to length scale, but with respect to particle number, too.  This is particularly relevant for interpreting the results in the language of halo finders, where the accuracy of halo properties is often considered as a function of particle number.

\subsection{Scale-Free Simulations}
\subsubsection{Definitions}
Scale-free simulations use the self-similar nature of gravity to probe the range of scales that a simulation faithfully reproduces.  An EdS ($\Omega_M=1$) background cosmology and a power-law power spectrum are introduced, leaving only one scale in the problem: the scale of onset of non-linearity.  At fixed time, this is given by a length scale, while at fixed length, this is given by a time scale.  This leads to the idea which lends scale-free simulations their utility: length and time may be used interchangeably as coordinates.  In other words, small-scale clustering at early times ought to be a rescaling of large-scale clustering at late times, giving a powerful tool to assess convergence.

The length and time scales are related as follows.  For a given power-law power spectrum with amplitude $A$ and spectral index $n_s$, the evolution under linear theory is given by
\begin{equation}
    P_L(k,a) = a^2Ak^{n_s}
\end{equation}
as a function of wavenumber $k$ and scale factor $a$.
The non-linear scale may be identified through the closely-related dimensionless power spectrum, given by
\begin{align}
    \Delta_L^2(k,a) &\equiv \frac{1}{2\pi^2} k^3 P_L(k,a) \nonumber \\
    &\propto a^2 k^{3+n}.
\end{align}
$\Delta_L^2$ gives the contribution to the variance in logarithmic intervals, and we have adopted proportionality since we are only interested in ratios of scales.  The self-similarity relation that yields constant $\Delta_L^2$ is therefore
\begin{align}
    \knl \propto a^{-2/(3+n)}.
\end{align}
Likewise, identifying $k\propto R^{-1}$ through the Fourier transform, we have
\begin{align}
    \Rnl \propto a^{2/(3+n)},
\end{align}
and identifying $M \propto R^3$ under the assumption of homogeneity, we have
\begin{align}
    \Mnl \propto a^{6/(3+n)}.
\end{align}

These scalings are expected to hold even in the deeply non-linear regime, $\Delta^2 \gg 1$.

The first output time of the simulation is chosen based on the epoch at which structures begin to form on small scales, as quantified by the small-scale variance.  Specifically, the variance of the overdensity with a spherical top-hat window $W_R(r)$ of radius $R$ is given by
\begin{equation}\label{eqn:sigma2}
    \sigma^2(R,a) = \int \Delta^2(k,a) \widetilde{W}^2_R(k) k^{-1} dk,
\end{equation}
where $\widetilde{W}^2_R(k)$ is the Fourier transform of $W_R(r)$.  The analytic solution to Eq.~\ref{eqn:sigma2} is presented in terms of gamma functions in \citetalias{Garrison+2021b}~(eq.~5).

The epoch of first output, $a_0$, is the scale factor at which 3-sigma fluctuations in the density field at radius $\ell = N^{1/3}/L$ (the mean interparticle spacing) reach the spherical overdensity collapse threshold of 1.68:
\begin{equation}\label{eqn:a0}
    \sigma(\ell,a_0) = 1.68/3 = 0.56.
\end{equation}

The normalization of the initial condition is also determined based on $\sigma$:
\begin{equation}
    \sigma(\ell,a_i) = 0.03.
\end{equation}

\subsubsection{Self-Similar Scaling of the $k$NN-PDF}
The self-similar scaling of the $k$NN-PDF may be determined as follows.  First, as the PDF is a function of distance $r$, it must be rescaled by a factor of $\Rnl$.  Second, we observe that $k$ is like a mass: the $k$-th nearest neighbor encompasses $k-1$ particles, by definition.  Therefore, it traces constant mass.  And as a mass, $k$ must be rescaled by \Mnl.

This introduces one extra wrinkle. $k$ only admits integer values, unlike the distance $r$, and must always be greater than 1.  \Mnl is therefore not arbitrary, but must be an even divisor of $k$.  Since time is labeled by \Mnl, the output epochs we wish to compare must fall on integer multiples of the first epoch.  In the simulation of Section \ref{sec:results}, the output epochs are related by a factor of $\sqrt{2}$ in \Mnl, so we compare every other epoch.  Similarly, the epoch of first analysis, where $k=1$, is not chosen to be the first epoch $a_0$ but rather an intermediate epoch that will trace scales relevant to halo formation.  Earlier epochs cannot be analyzed simultaneously because they would have $k<1$.

\section{Application to a $\lowercase{n_s}=-2$ Simulation}\label{sec:results}
\subsection{Overview}
We measure the $k$NN-PDF at multiple epochs of an $n_s=-2$ scale-free simulation, apply the self-similar rescaling, and assess the convergence as a function of epoch and length scale, or equivalently particle number and overdensity.  We divide our analysis into two regimes: high-density (halos) and low-density (voids).  We first discuss the details of the simulation, then turn to the analysis.

\subsection{Simulation}
The simulation used in this work is a $N=1024^3$ particle simulation, first presented in \citetalias{Joyce+2021}, run with the \Abacus $N$-body code \citep{Garrison+2021a}.  \Abacus offers high force accuracy and uses many global time steps, minimizing integration errors in the particle trajectories\footnote{A working title of this paper was ``Good Forces Make Good Neighbors''}.  Using a high-order multipole method on a static mesh to solve the far-field force, and an analytically disjoint near-field force calculation accelerated by GPUs, \Abacus achieves simultaneous high performance and high accuracy.

This simulation employed a spline softening fixed in comoving coordinates with a Plummer-equivalent length of $\epsilon=\ell/30$, and a time step parameter of $\eta_\mathrm{acc}=0.15$.  Variations in these choices were explored in \citetalias{Garrison+2021b}, with the time step found to be conservative, and diminishing returns found for reducing the softening below $\ell/30$.

The simulation produced full particle outputs at 38 epochs, logarithmically spaced in scale factor (and therefore \Mnl).  Specifically, the outputs were spaced by a factor of $\sqrt{2}$ in \Mnl.  Since we seek to scale $k$ by an integer value in order to compute the $k$NN-PDF at multiple epochs, every other output was used, yielding a factor of 2 in \Mnl between each epoch.  We note that while the epoch of first output is $a_0$ (Eq.~\ref{eqn:a0}), this is not the first epoch of analysis, which was chosen to be $\Mnl = 512$, or $a/a_0 = 2.83$, so that the small-scale, high-density regions would be probed at late times.

\subsection{Analysis of the $k$NN-PDF}
\subsubsection{Measurement and Rescaling}

\begin{figure}[t]
    \centering
    \includegraphics[width=\columnwidth]{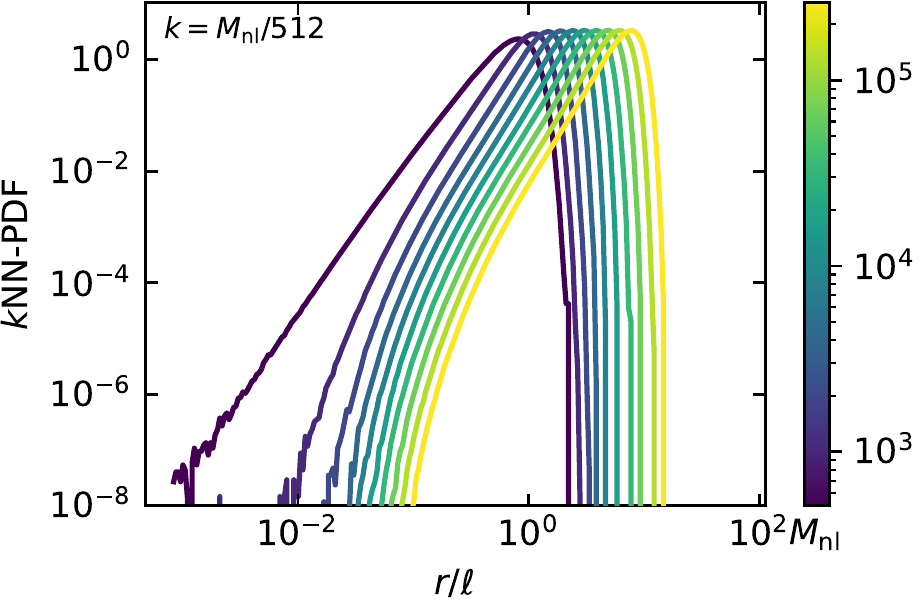}
    \caption{The probability distribution function of finding the $k$-th neighbor at distance $r$, expressed in units of the mean interparticle spacing $\ell$.  Each line corresponds to a different epoch, labeled by \Mnl (colorbar).  The $k$ is a function of epoch and is given by $k=\Mnl/512$, spanning $k=1$ at the earliest epoch to $k=512$ at the latest.}
    \label{fig:pdf}
\end{figure}

We measure the $k$NN-PDF on 10 epochs between $\Mnl = 512$ and $262144$, or $k=1$ and $512$ using $N_R=4\times10^8$ random points. Rather than a uniform random distribution of points, we design an importance sampling function that upweights the low- and high-density regions of the simulation to reduce the noise in the tails of the distribution (Appendix \ref{sec:sampling}).

The measurements are shown in Figure~\ref{fig:pdf}.  Each line represents a different epoch, with an overall rightward shift occurring towards later epochs, as $k$ increases.  The overall shape is very roughly that of a Gaussian in log-space, but with a long tail to low $r$---the result of clustering, as some regions of space will have many neighbors packed closely together.  The peak of the distribution occurs approximately at the mean density: consider that the first epoch (darkest line), with $k=1$, peaks at $r/\ell=1$.  The last epoch (lightest line), with $k=512$, peaks at $r/\ell=512^{1/3}=8$.

\begin{figure}[t]
    \centering
    \includegraphics[width=\columnwidth]{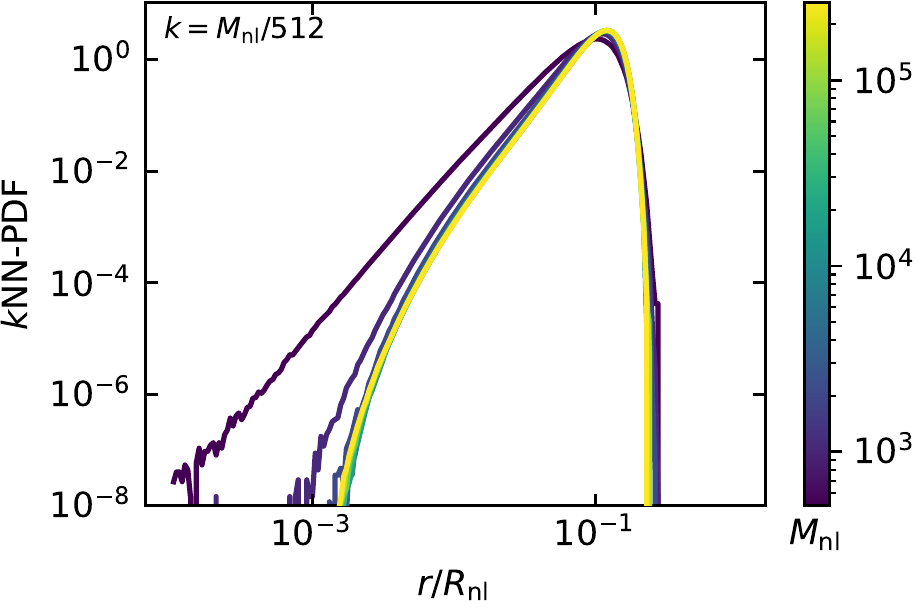}
    \caption{Same as Fig.~\ref{fig:pdf}, but in rescaled units of $r/\Rnl$.  The PDF is seen to exhibit approximate self-similarity, with the lines superimposing.}
    \label{fig:pdf_rescaled}
\end{figure}

The self-similar rescaling of the PDF is shown in Figure~\ref{fig:pdf_rescaled}.  Immediately one sees that the PDFs do rescale self-similarly to a good approximation---that is, they stack---but with notable outliers at the earliest epochs (smallest $k$).  Elsewhere, the agreement is good, although both axes span many orders of magnitude, so small differences are difficult to discern.

\begin{figure}[t]
    \centering
    \includegraphics[width=\columnwidth]{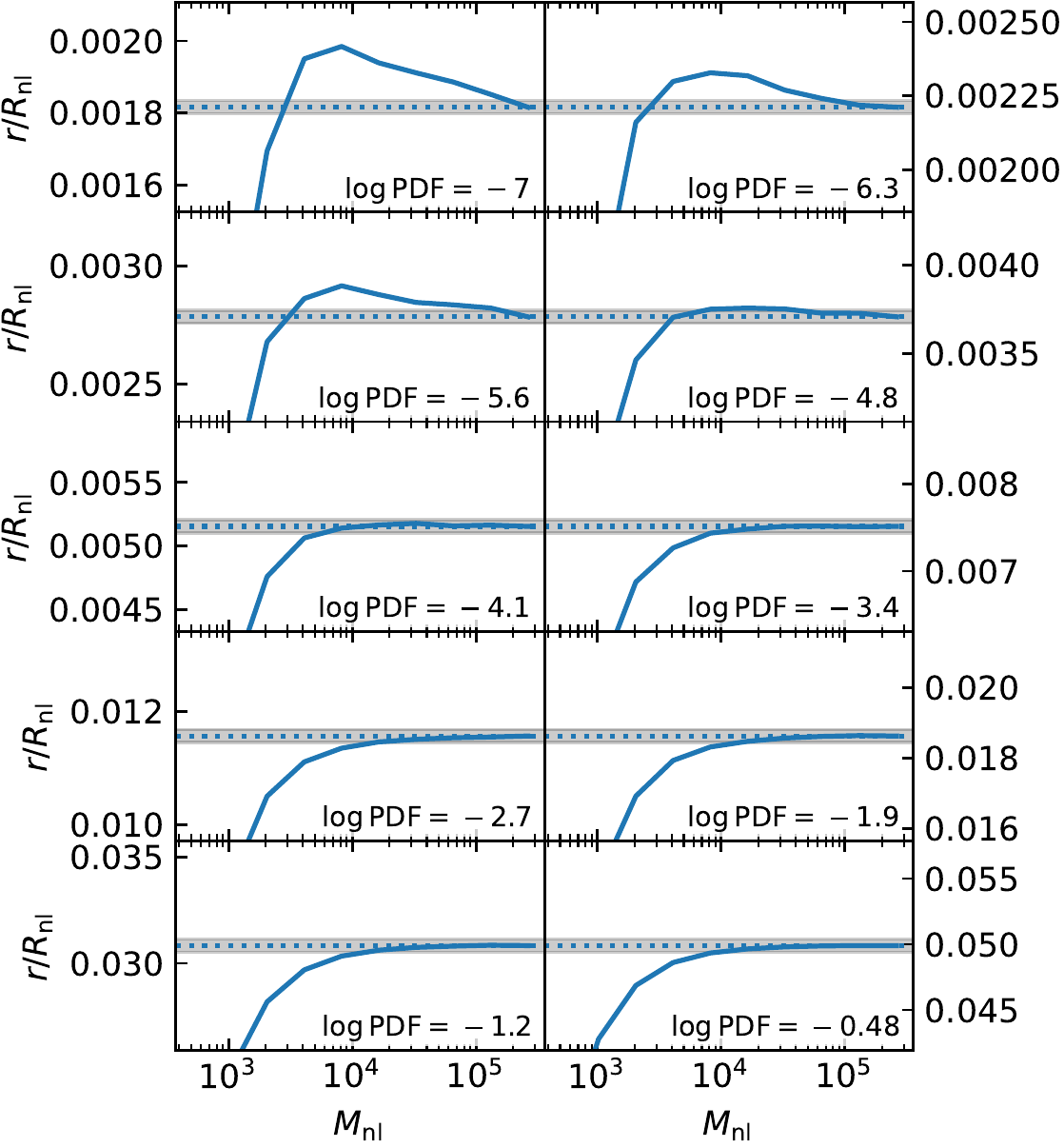}
    \caption{Convergence of $r/\Rnl$ as a function of epoch, for multiple values of the PDF (panels).  Each panel may be thought of as a horizontal slice through the lines in Fig.~\ref{fig:pdf_rescaled}, considering only the high-density tail of the PDF (small $r$).}
    \label{fig:slices_highdens}
\end{figure}

To make a quantitative analysis, we make multiple ``slices'' of the PDFs horizontally, and plot the $r/\Rnl$ value where each slice intersects the PDF for each epoch in Figure~\ref{fig:slices_highdens}.  We slice horizontally because the steepness of the PDF makes vertical slices difficult to assess, especially in the large $r$, low-density tail.  For the moment, however, we will focus solely on horizontal slices of the high-density, small $r$ tail, leftward of the peak.

\subsubsection{High-Density Regime}\label{sec:highdens}

Fig.~\ref{fig:slices_highdens} shows 10 such slices, one in each panel, ranging from $\log_{10}(\mathrm{PDF})=-7$ to $-0.48$ (in detail, we make 30 such slices, but only 10 are plotted here).  The $x$-axis labels the epoch by \Mnl, while the plotted quantity is the value of the $x$-axis in Fig.~\ref{fig:pdf_rescaled} intersected by the horizontal slices.  Flat lines indicate constant value, unchanging over epoch; i.e., self-similarity.  This recalls the analysis of \citetalias{Garrison+2021b}, in which flat lines---constant correlation function amplitude---indicated self-similarity.

At early times (leftward in each panel), we see a lack of convergence (steep lines), but this flattens towards convergence in all but the smallest values of $\log(PDF)$ (smallest $r/\Rnl$ values).  A region of $\pm1\%$ is shown as a shaded band in this figure; epochs that fall within this band may be declared to exhibit self-similarity at the 1\% level (with the exception of those epochs that momentarily cross through the band).

\begin{figure}[t]
    \centering
    \includegraphics[width=\columnwidth]{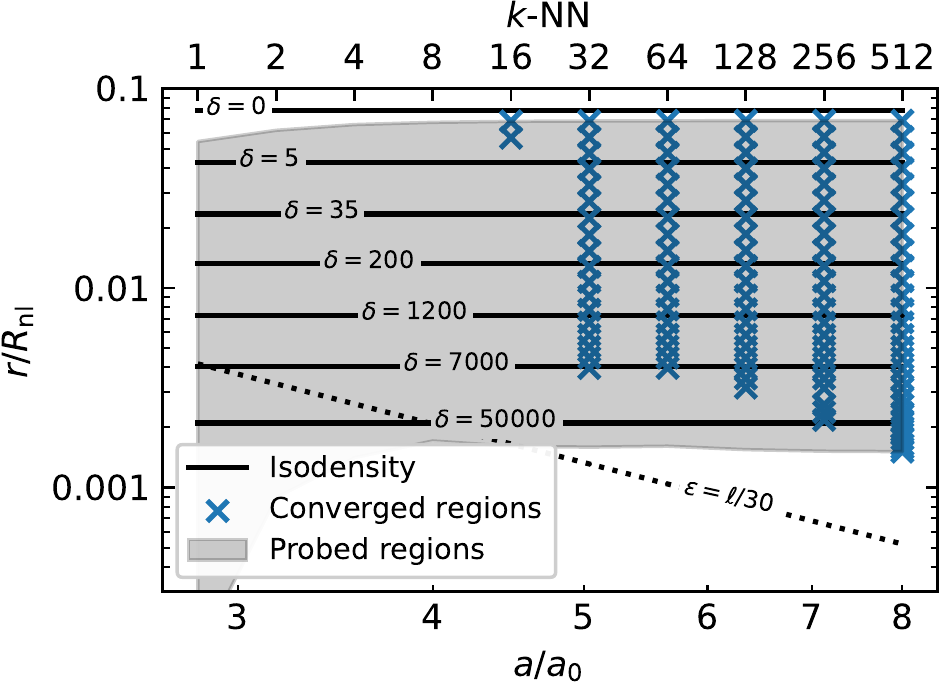}
    \caption{The map of converged mass and length scales, as determined from Fig.~\ref{fig:slices_highdens}.  Each blue cross is an epoch whose $k$NN-PDF is converged to self-similar solution at that $r/\Rnl$; the gray shaded region is the parameter space covered by the analysis.  The top axis labels the number of particles at that epoch, and the solid lines are isodensity contours.  Therefore, at 16 particles and fewer, we hardly find convergence at any density, while at 32 particles and greater, the range of converged densities increases tremendously to $\delta>7000$, well within halo cores.}
    \label{fig:convergence_highdens_rescaled}
\end{figure}

\begin{figure}[t]
    \centering
    \includegraphics[width=\columnwidth]{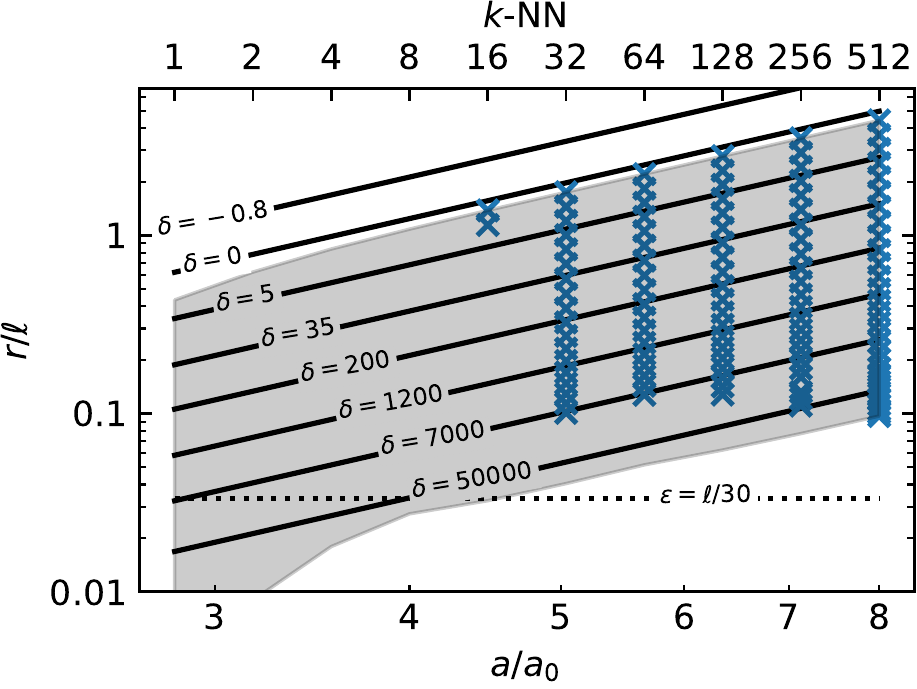}
    \caption{Same as \ref{fig:convergence_highdens_rescaled}, but plotting $r/\ell$, where $\ell$ is the (non-rescaled) mean interparticle spacing.}
    \label{fig:convergence_highdens}
\end{figure}

In Figure~\ref{fig:convergence_highdens_rescaled}, we mark those epochs that fall within the 1\% region with blue crosses.  The $x$-axis is epoch, now labeled by $a/a_0$ on the bottom axis, or $k$ on the top axis.  The $y$-axis is the $r/\Rnl$ value to which the $k$NN-PDF converges for each epoch---the dashed lines in Fig.~\ref{fig:slices_highdens}.  The gray shaded region indicates the $r/\Rnl$ values this analysis is sensitive to, as a function of epoch---the gray regions without blue crosses are where we tested for convergence, but did not find it.

We can interpret Figure~\ref{fig:convergence_highdens_rescaled} as follows.  For $k\le8$ (8 particles or fewer), we do not find convergence at any time or length scale.  At 16 particles, we find convergence over a narrow range of length scales, which rapidly expands to smaller radii at 32 particles.  From 64 particles to 512 particles, the improvement is less dramatic, but steadily extends to smaller length scales.

We can connect this analysis to the language of spherical overdensity, $\delta = \rho/\overline{\rho}-1$.  Because the $k$NN returns the radius of a sphere that encompasses a fixed number of particles $k$, we may immediately convert such a radius into an overdensity.  Indeed, the self-similar rescaling of $k$ with $\Mnl$ exactly probes an isodensity contour for fixed $r/\Rnl$, which is why we expect the $k$NN-PDF to given identical answers across epoch in the first place.  These isodensity contours are plotted as solid lines in Fig.~\ref{fig:convergence_highdens_rescaled}.

Now interpreting the convergence relative to overdensity, we see that the narrow convergence at 16 particles occurs at quasi-linear densities, from $\delta\sim0$ to $3$.  At 32 particles, the upper limit increases by three orders of magnitude to $\delta=7000$, typical of halo cores.  Therefore, we may tentatively conclude that halos of 32 particles and above are converged in a spherical overdensity sense, at least only considering the mass interior and not the details of internal structure.

Fig.~\ref{fig:convergence_highdens} shows the same information as Fig.~\ref{fig:convergence_highdens_rescaled}, except with the $y$-axis now in units of $r/\ell$ instead of self-similar units.  Notably, one may now readily locate the mean interparticle spacing at $r/\ell=1$, and see that no particular damage is done to the convergence of the $k$NN-PDF by any ``memory'' of the initial particle lattice in the late-time particle distribution, which would peak at this location.  This is consistent with \citetalias{Garrison+2021b}, in which the lattice memory was apparent at early epochs but erased effectively as the correlation length exceeded $r=\ell$.  We see that the minimum resolved comoving length scale is about $\ell/10$, with modest decrease at later epochs.

\subsubsection{Low-Density Regime}

\begin{figure}[t]
    \centering
    \includegraphics[width=\columnwidth]{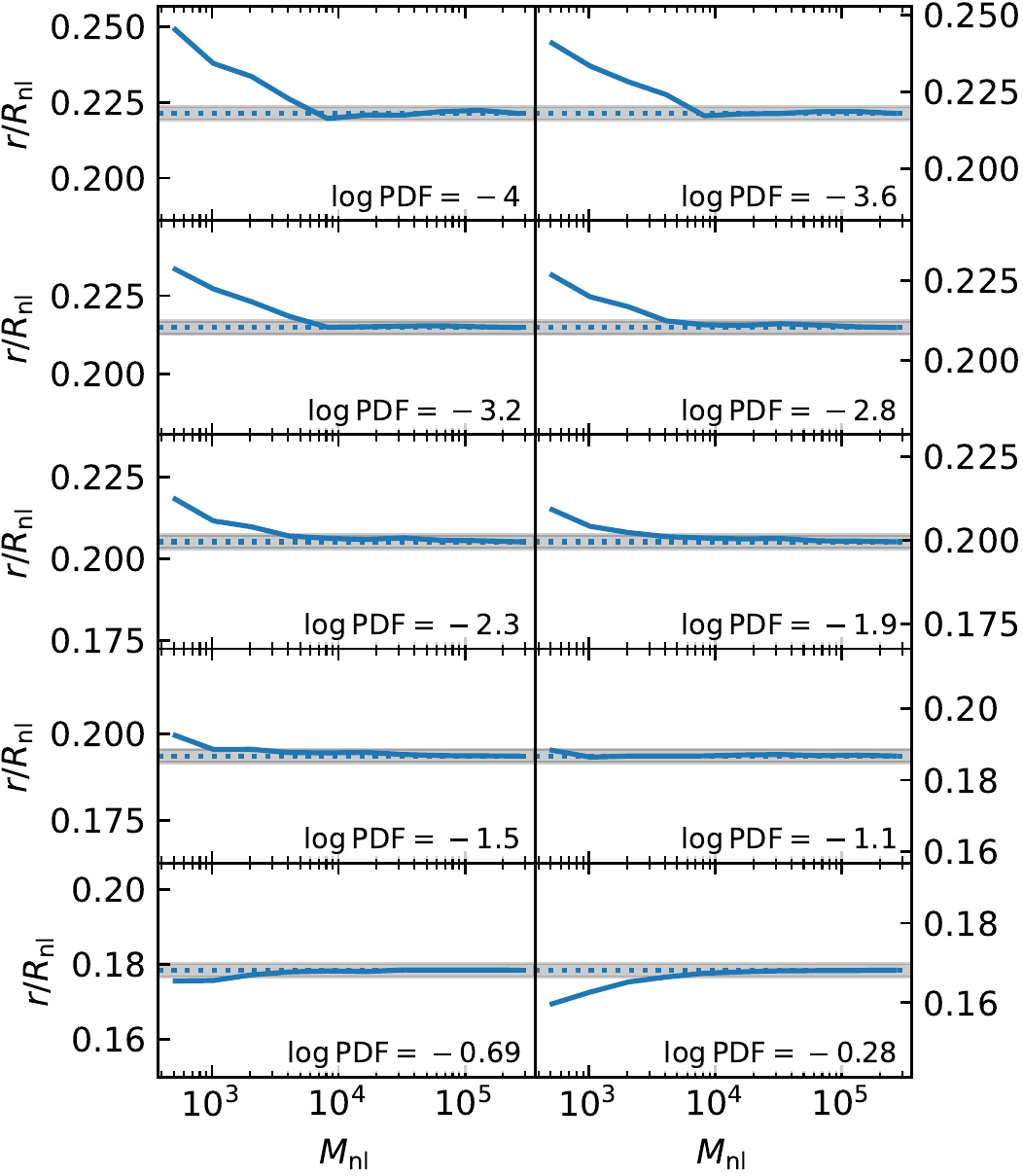}
    \caption{Same as Fig.~\ref{fig:slices_highdens}, but for the low-density tail of the $k$NN-PDF, taking horizontal slices through the large $r$ branch of Fig.~\ref{fig:pdf_rescaled}.}
    \label{fig:slices_lowdens}
\end{figure}

\begin{figure}[t]
    \centering
    \includegraphics[width=\columnwidth]{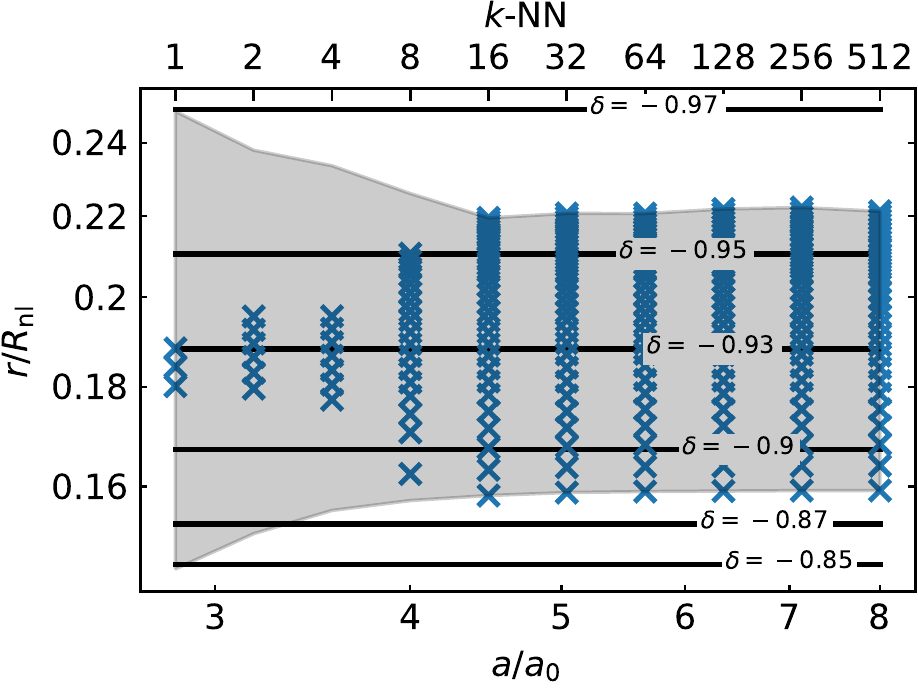}
    \caption{The map of converged mass and length scales in the low-density regime, as determined from Fig.~\ref{fig:slices_lowdens}.  Plotting elements are as in Fig.~\ref{fig:convergence_highdens_rescaled}; blue crosses indicate convergence.  We find evidence for convergence in the whole range of probed densities at 16 particles and greater, although the range of densities, $\delta=-0.88$ to $-0.955$, is narrow.
    }
    \label{fig:convergence_lowdens_rescaled}
\end{figure}

\begin{figure}[t]
    \centering
    \includegraphics[width=\columnwidth]{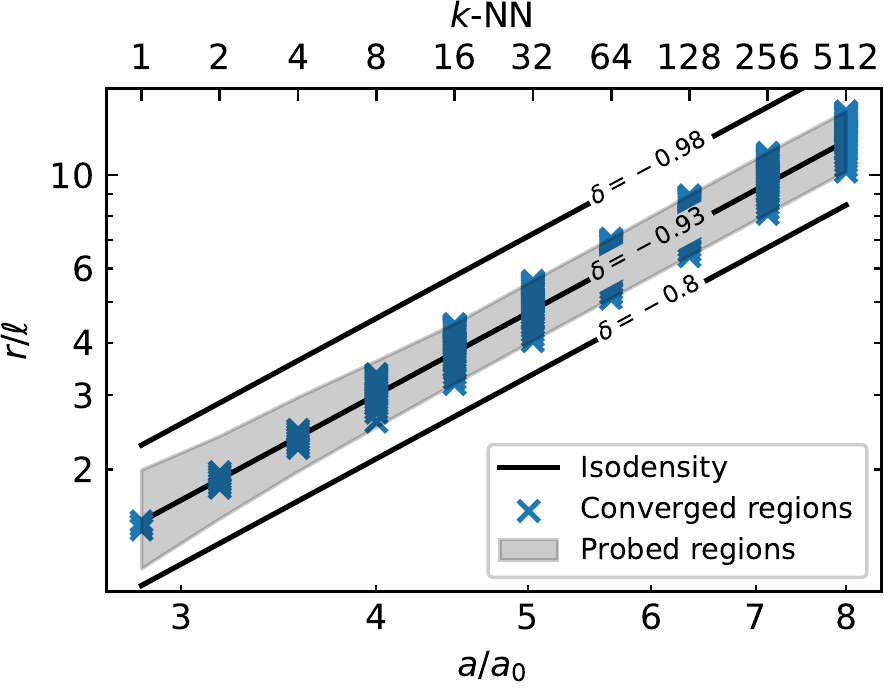}
    \caption{Same as \ref{fig:convergence_lowdens_rescaled}, but plotting $r/\ell$, where $\ell$ is the (non-rescaled) mean interparticle spacing.  Since $r/\ell=1024$ is the box scale, the largest measured $r/\ell$ values reach about 1/50th of the box scale.}
    \label{fig:convergence_lowdens}
\end{figure}

Rightward of the peak in the $k$NN-PDF in Fig.~\ref{fig:pdf_rescaled}, we have the low-density tail, which probes voids where the density is $\delta<0$.  The PDF is particularly steep here, so it does not cover a very wide range of densities, but we nonetheless repeat the same procedure of the previous section: make horizontal slices of the PDF (Fig.~\ref{fig:slices_lowdens}), determine the regions converged to within 1\%, and map them as a function of number of particles and overdensity (Figs.~\ref{fig:convergence_lowdens_rescaled} \& \ref{fig:convergence_lowdens}).

Examining Figure~\ref{fig:convergence_lowdens_rescaled}, we see qualitatively different behavior from the high-density case.  For voids with 4 or fewer particles, there is only a narrow $\delta$ range of convergence, around $\delta=-0.92$.  Examining Figure~\ref{fig:convergence_lowdens_rescaled}, we see that this range of particles underestimates the density at low density, and overestimates the density at high density.  In other words, it ``swings'' from too-high to too-low $r/\Rnl$, and the ``convergence'' occurs when it passes through the midpoint.  It is therefore likely that this convergence is unphysical, but the range of densities it covers is so narrow as to be negligible.  However, this becomes less clear as the range of densities expands at 8 particles, so we choose to present the whole set of measurements.

At 16 particles and greater, the whole range of covered densities is converged, from $\delta=-0.88$ to $\delta=-0.955$.  Of course, this is a small range of densities, but it is nonetheless interesting that 16-particle voids agree with 512-particle voids of the same density at later times.

In Figure \ref{fig:convergence_lowdens}, we see the same information as \ref{fig:convergence_lowdens_rescaled} but with the $y$-axis in units of $r/\ell$ rather than scale-free units.  Here, we see that the comoving size of these voids at late times is quite large, greater than 1/100th of the box scale in the last epoch.  While this would not be considered very large in an cosmological $\Lambda$CDM simulation, a scale-free simulation with a red spectrum has significantly more large-scale power because its power spectrum does not have a turnover at the peak of matter-radiation equality.  Scale-free simulations are therefore more sensitive to finite box size effects; indeed, hints of such effects were present in the correlation function at $L/100$ in \citetalias{Garrison+2021b}.  However, we see no such effects here.  Possibly larger scales or different overdensities would be more sensitive, or perhaps the nature of the $k$NN-PDF as an ``interior mass'' measure mixes scales more efficiently than the 2PCF.

The range of densities in the analysis could be extended if one is willing to analyze the $k$NN-PDF near the peak.  Because the PDF becomes flat here, this ``horizontal slice'' procedue is not appropriate.  A complementary analysis could be done using vertical slices in a restricted range, although interpreting the results on equal footing with the horizontal slices might not be straightforward, as error tolerances likely have different interpretation.

\subsection{Softening Length \& Comparison with Two-Point Analysis}

\begin{figure}[t]
    \centering
    \includegraphics[width=\columnwidth]{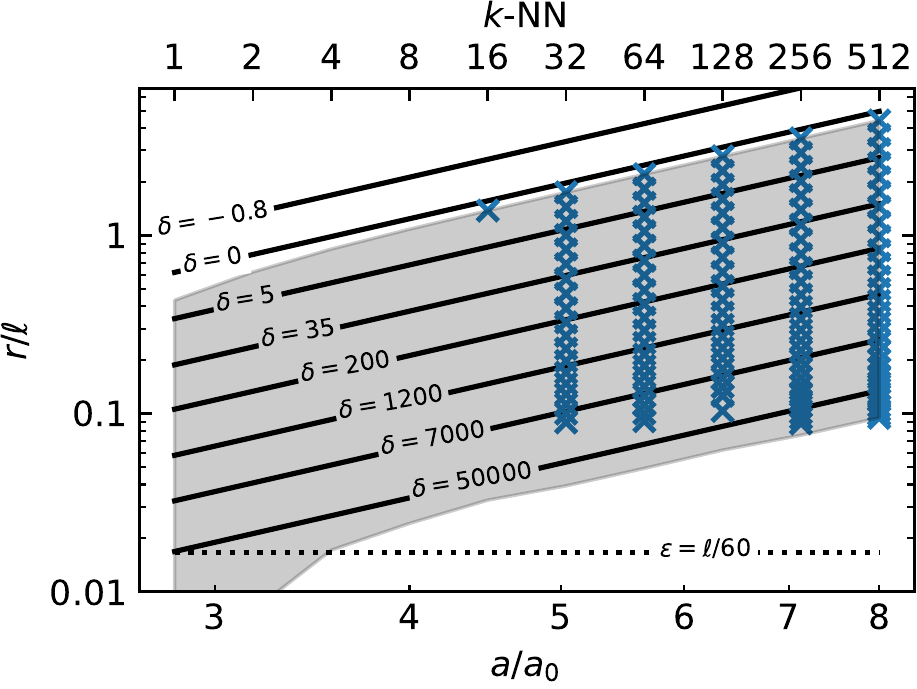}
    \caption{Same as Fig.~\ref{fig:convergence_highdens}, but for a simulation with half the softening length ($\epsilon=\ell/60$).  In the $k=32$ to $256$ regime, we see that the convergence extends to smaller length scales (higher $\delta$).  Notably, the smaller softening does not extend the reach of convergence to fewer particles (smaller $k$), at least at the factor-of-two mass granularity of this analysis.
    }
    \label{fig:convergence_highdens_halfeps}
\end{figure}

We may compare our results to the two-point analysis of \citetalias{Joyce+2021} and \citetalias{Garrison+2021b}, at least in the high-density regime common to both analyses (Sec.~\ref{sec:highdens}).  In this work as in theirs, the same qualitative picture emerges of convergence propagating from large to small scales as the simulation progresses.  In Fig.~\ref{fig:convergence_highdens}, we see that the smallest resolved comoving length scale is about $r=\ell/10$; comparing with \citetalias{Garrison+2021b}'s figure 9, we find remarkable agreement.  The progression from large to small scales is more evident in that work, proceeding from $0.15\ell$ to $0.1\ell$ over the same range of epochs, although 128 to 512 particles shows some improvement, too.

\citetalias{Garrison+2021b} show that $\epsilon=1/30$ roughly matches the limit set by the mass resolution, but that halving the softening length to $\ell/60$ still produces a small gain---far from a factor of two, but still measurable.  To test this in the $k$NN statistic, we repeat the analysis of Section~\ref{sec:highdens} for an identical simulation with half the softening length.  The result is shown in Fig.~\ref{fig:convergence_highdens_halfeps}, where we indeed find that the resolution improves, by about 10\% to 30\% for 32 to 256 particles.  We cannot measure any improvement at 512 particles, because it is already saturated to the highest density in our analysis, and the steepness of the PDF precludes probing higher densities.

Notably, halving the softening length does not increase the range of resolved masses---32 particles remains resolved, and 16 particles unresolved.  However, this analysis has only a factor-of-two granularity in mass, so it is yet possible that there is some gain between 16 and 32 particles.

\section{Conclusions}\label{sec:conclusions}
We have measured the $k$-nearest-neighbor probability distribution function on a scale-free cosmological $N$-body simulation with spectral index $n_s=-2$ and used it to determine the length and mass scales over which the simulation is converged.  This extends the analysis of \citetalias{Joyce+2021} and \citetalias{Garrison+2021b} because the $k$NN-PDF encodes information from higher-order correlations of the density field, beyond the two-point function.  By applying a self-similar rescaling of both $k$ and the length scale, we find that the $k$NN-PDF exhibits broad self-similarity.  To quantify this, we divide our analysis into two parts: the high-density regime (halos), and the low-density regime (voids).

In the high-density regime, we map convergence in terms of number of particles $k$, and overdensity $\delta$ (Fig.~\ref{fig:convergence_highdens}).  For 8 particles and fewer, we find no evidence for convergence at any density between $\delta=0$ and $10^5$.  At 16 particles, we find a narrow range of convergence near $\delta=0$, but this range quickly expands with more particles.  At 32 particles, we see convergence for $\delta<7000$, well within halo cores.  This improves to $\delta=10^5$ by 512 particles.  The resolved comoving length scale is $\ell/10$ stable within 30\% from 32 to 512 particles.

Comparing these results to the two-point analysis of \citetalias{Joyce+2021} and \citetalias{Garrison+2021b}, we find excellent agreement, with these works suggesting a minimum resolved length scale of $0.15\ell$ to $0.1\ell$ over the same range of epochs.  We repeat our $k$NN analysis with a simulation of half the softening length, and find modest improvements of 10\% to 30\% in resolved length scale---slightly greater than in the two-point work, but not a factor of two.  This is consistent with $\epsilon=\ell/30$ yielding most of the possible resolution at a given particle mass.

In the low-density regime, our analysis is sensitive to a smaller range of densities, from $\delta=-0.88$ to $-0.955$.  Still, in this regime, we find weak evidence for convergence at 8 particles and fewer, and stronger evidence at 16 particles and greater.  The analysis reaches about 1/100th of the box size, but we do not observe the large-scale effects that we attributed to finite box-size in the two-point analysis of \citetalias{Garrison+2021b}.

The results of both the halo and void analysis may be seen as surprisingly optimistic; 32-particle halos are not considered particularly robust, and likewise it is hard to imagine that the dynamics of 16-particle voids are well-sampled, yet the $k$NN-PDF considers both well-converged.  Part of the answer may lie in the effective smoothing the $k$NN imposes on the density field.  That is, it says nothing about the internal structure of 64-particle halos or any property other than their spherically-averaged mass.  The steepness of the PDF may play a role too, especially in the low-density regime.  Horizontal slicing of the PDF was chosen to mitigate this effect, but voids are fundamentally already confined to a narrower range of densities ($\delta=-1$ to $0$) than their high-density counterparts.  Therefore, the range of radii voids produce at fixed density is narrower, giving the $k$NN, which measures radii, less of a lever-arm.

Of course, we have only explored a small range of void densities in this work, and it is possible an extended range of densities or a more sophisticated analysis that mitigates the steepness of the $k$NN-PDF have greater discerning power.

\begin{acknowledgements}
The authors would like to thank Arka Banerjee for productive discussions, and Michael Joyce and Sara Maleubre for their contributions to the two-point analysis.

Abacus development has been supported by NSF AST-1313285 and DOE-SC0013718, as well as by Harvard University startup funds.
DJE is supported in part as a Simons Foundation investigator. 
LHG is supported by the Center for Computational Astrophysics at the Flatiron Institute, which is supported by the Simons Foundation.
TA is supported in part by the U.S. Department of Energy, and the U.S. Department of Energy SLAC Contract No. DE-AC02-76SF00515.
\end{acknowledgements}

\section*{Data availability}
Clustering measurements and other summary statistics used in this work are available upon request.  The underlying simulation data is much larger, but may also be made available upon reasonable request.

\bibliography{biblio}{}
\bibliographystyle{aasjournal}

\appendix
\section{Importance Sampling for the $\lowercase{k}$NN-PDF}\label{sec:sampling}
Measurement of the $\lowercase{k}$NN-PDF usually involves generating a set of $N_R$ uniform random points, querying the data points for the the $k$-th nearest distance to each random point (often with a $k$d-tree), and estimating the probability distribution function from those distances (often with a histogram or interpolation).  The finite size of $N_R$ introduces some noise in the estimation of the PDF, especially at the tails of the distribution.  Consider that finding the low-distance tail requires a bullseye of a high-density region like a halo, which occupies relatively little volume.  Similarly, finding the large-distance tail requires landing in the middle of a void.

Noise in the estimation of the PDF can be reduced with importance sampling, in which the random points are generated with greater probability in high- and low-density regions and are assigned weights in inverse proportion to their probability.  Specifically, we design the following probability function $P(\mathbf{x})$:
\begin{equation}\label{eqn:importance}
    P(\mathbf{x}) \propto \min((\rho(\mathbf{x})/\overline{\rho} + 0.01, 10).
\end{equation}
This weights the probability in proportion to the density so that halos are densely sampled, but with a uniform background level that ensures voids are sampled well enough to find their centers.  The density peaks are clipped to keep the dynamic range of weights to a factor of 1000.  Each random particle $i$ is then assigned weight $w_i = 1/P(\mathbf{x}_i)$ in the computation of the PDF.

To estimate $\rho(\mathbf{x})$, we generate a density grid using triangle-shaped cloud (TSC) mass assignment on a $4096^3$ mesh at the first epoch, scaled to larger cell sizes in proportion to $\Rnl$ at later epochs.  This relatively fine mesh was found to help effectively localize small halos at early times, and indeed yields a cell size of only $\ell/4$, while the most extreme tails we seek are at $\ell/100$.

Generating random points according to $P(\mathbf{x})$ can either be accomplished with rejection sampling or Poisson draws.  The rejection sampling method throws random points in the volume and accepts them in proportion to the probability of the cell in which they land.  The Poisson draw method goes cell-by-cell, drawing a Poisson value for the occupation number of that cell, and then generates that many uniform random particles within the cell.  We find the Poisson method more robust, as it is insensitive to the details of $P(\mathbf{x})$.  The rejection sampling method can suffer from catastrophically low acceptance rates for aggressive $P(\mathbf{x})$.  Both methods parallelize well.  We use Numba for the implementation in this work \citep{10.1145/2833157.2833162}.

The entire $k$NN-PDF algorithm is thus the following:
\begin{enumerate}[(i)]
    \item load the data points,
    \item generate a TSC density field,
    \item generate $N_R$ random points according to the density (Eq.~\ref{eqn:importance}) and record their inverse probabilities as weights,
    \item construct a $k$d-tree from the data points,
    \item query the tree with the $N_R$ randoms for each's $k$-th neighbor distance,
    \item histogram the distances, with each distance using the weight of the corresponding random.
\end{enumerate}

This produces about 3 GB of data per epoch when measured with $N_R=4\times10^8$.  In our implementation, \textsc{SciPy}'s single-threaded $k$d-tree construction is the slowest step, taking about 500 seconds for $N=1024^3$.  Using 128 cores, generating importance-sampled randoms takes 180 seconds for a $4096^3$ mesh, and the tree query 80 seconds.

\begin{figure}
    \centering
    \includegraphics[width=\columnwidth]{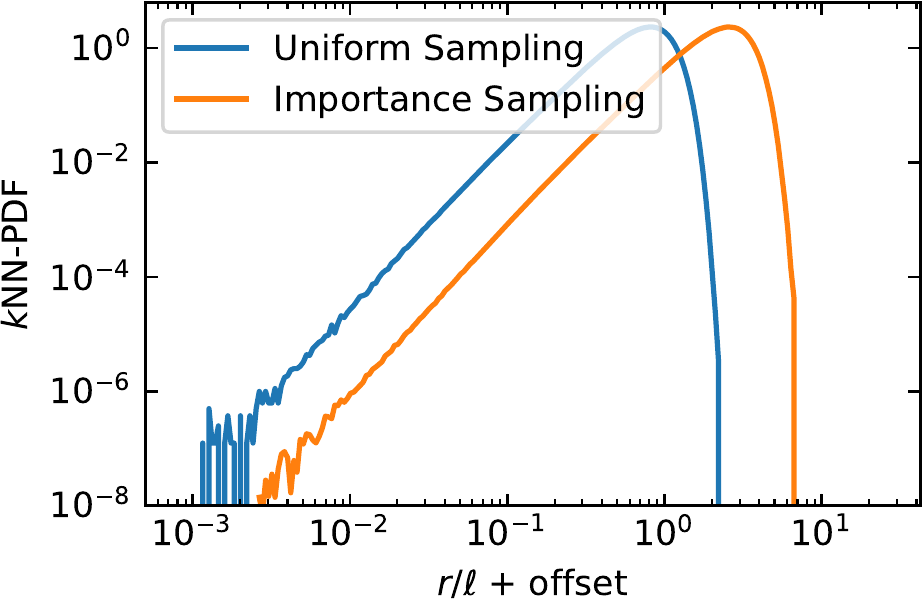}
    \caption{The $k$NN-PDF for the earliest epoch from Fig.~\ref{fig:pdf}, with and without importance sampling.  The importance-sampled result (\textit{orange line}) is seen to reduce the noise in the low $r$ tail.  This line has an arbitrary $x$-offset applied to aid the visual comparison.}
    \label{fig:importance_pdf}
\end{figure}

\end{document}